\documentclass[aps, twocolumn, showpacs, amsmath]{revtex4}
\usepackage{dcolumn}% Align table columns on decimal point
\usepackage{bm}% bold math
\usepackage{graphicx}% Include figure files
\usepackage{lipsum}
\usepackage{color}
\usepackage{subfigure}
\usepackage{hyperref}
\usepackage{latexsym}
\usepackage{amsthm}
\usepackage{amssymb}
\usepackage{braket}
\DeclareGraphicsExtensions{.jpg,.pdf, .mps, .png, .eps, .ps, .EPS,.gif}

\begin{document}

\def\be{\begin{equation}}
\def\ee{\end{equation}}

\def\bc{\begin{center}}
\def\ec{\end{center}}
\def\bea{\begin{eqnarray}}
\def\eea{\end{eqnarray}}
\newcommand{\avg}[1]{\langle{#1}\rangle}
\newcommand{\Avg}[1]{\left\langle{#1}\right\rangle}

\def\ie{\textit{i.\,e.,}}
\def\etal{\textit{et al.}}
\def\m{\vec{m}}
\def\G{\mathcal{G}}

\newcommand{\red}[1]{{\bf\color{green}#1}}
\newcommand{\eq}[1]{Eq.\ (\ref{eq:#1})}

% Figures 
% \includegraphics[width=2.5  in]{picL064pcfor53894043.eps} %{picpc} 
%   \includegraphics[width=2.5  in]{picL064pc2for53894043.eps}%//{picpc2} 
%  \includegraphics[width=2.5  in]{picL064pc2-7-1.eps} %{picpc2}   Fig 2a
%   \includegraphics[width=2.5  in]{picL064pc-7-1.eps}  %{picpc}   Fig 2b
%   \includegraphics[width=3.5  in]{mannaP2p} 
%  \includegraphics[width=3.5  in]{ScalingPlotManna}           Fig. 5
%   \includegraphics[width=3 in]{FigMannaPk.pdf} 
%   \includegraphics[width=3 in]{manna081219figurek1.eps}
 %   \includegraphics[width=3 in]{manna081219figurek2.eps}
 %   \includegraphics[width=3 in]{manna081219figurek3.eps}
 %   \includegraphics[width=3 in]{manna081219figurek4.eps}
                  
%Fig 3   Prpc.eps

%\newcommand{\rev}[1]{{\color{red}#1}}  %%% use this version to show revisions in red
\newcommand{\rev}[1]{#1}    %hide revisions

\title{Universal correlations in percolation}

\author{Robert M. Ziff}
\affiliation{Center for the Study of Complex Systems and Department of Chemical Engineering, University of Michigan, Ann Arbor, Michigan 48109-2800, USA}

\begin{abstract}

We discuss correlations in percolation and recent contributions of 
Xiaojun Tan, Youjin Deng and Jesper Lykke Jacobsen in their paper ``N-cluster correlations in four- and five-dimensional percolation," {\it Frontiers of Physics} {\bf 15} 41501 (July, 2020).  This is a View and Perspective in {\it Frontiers of Physics}.

\end{abstract}

\maketitle

%\section{Introduction}
%\label{sec:introduction}

Percolation is the process where connections from link to link in random systems leads to clustering and long-range connectivity.  A standard model of percolation is a regular lattice, with the links being nearest-neighbor bonds that are randomly occupied (bond percolation), or all links exist and random sites (vertices) are occupied (site percolation).   At a critical occupancy fraction $p_c$ of sites or bonds, the mean cluster size in an infinite system first becomes infinite, and many properties become universal for a given dimensionality.  The characteristics of the universal behavior has been the focus of much research for decades \cite{Cardy92,SmirnovWerner01,SaleurDuplantier87,StaufferAharony1994,LanglandsEtAl92}.

Correlations  follow universal power-law behavior at the critical point.  Define $P(r_1,r_2)$ as the probability that two lattice points a distance $r=|r_1-r_2|$ apart belong to the same cluster.  For large $r$, at the critical threshold, $P$ is expected to behave as 
\be
P(r_1,r_2) \sim r^{-2(d-d_f)}
\label{eq:P2}
\ee
where $d$ is the dimensionality and $d_f$ is the fractal dimension, with the values $d_f = 91/48$ in two dimensions (2D), 2.5229 in 3D, etc.  While these exponent values are expected to be the same for all percolation systems at criticality (i.e., universal), the coefficient in \eq{P2} cannot be universal because it depends upon the lattice-level behavior and thus the specific lattice and percolation type (site or bond). Furthermore, if one takes a continuum limit of percolation where the lattice mesh size goes to zero, the probability that a point belongs to a given cluster goes to zero, because of the fractal nature of critical percolation clusters.   Universality in that limit can recovered by looking at properties that do not depend upon lattice-level properties, such as the probability of crossing from one segment of the boundary of a finite system to another.   For correlation functions, one can recover universality by looking at certain ratios of functions.  If we define $P_\varepsilon(r_1,r_2)$ as the probability that there is connectivity between the surfaces of two spheres of radius $\varepsilon$ around each of the two points $r_1$ and $r_2$, then this quantity survives in the limit that the mesh goes to zero, and will be proportional to the product of the two surface areas $(\varepsilon^{d-1})^2$.  By taking ratios of correlations functions, one can get quantities whose factors cancel out, and survive in the limit $\varepsilon \to 0$, and many of these ratios are  accessible by conformal theory methods.  For example, the ratio
\be
R = \frac{P(r_1,r_2,r_3)}{\sqrt{P(r_1,r_2) P(r_1,r_3) P(r_2,r_3)}}
\label{eq:R}
\ee
where $P(r_1,r_2,r_3)$ is the probability that three far-separated points connect together in a single cluster, has a universal value of $2^{7/2} 
   3^{-3/4} \pi^{5/2} \Gamma(1/3)^{-9/2} \approx 1.0299$ \cite{SimmonsZiffKleban09} 
   when two of the points are on the boundary, and $1.02201$
   when all three points are internal  \cite{SimmonsZiffKleban09,DelfinoViti10}.

In Ref.\ \cite{VasseurJacobsenSaleur12}, Vasseur, Jacobsen and Saleur introduced the idea of looking at correlations between two groups of neighboring points, the groups separated a long distance from each other, and looked at different possible connections.  They show that certain combinations of these correlations have power-law behavior, with exponents in 2D that can be predicted theoretically, and also identified a ratio $(F(r)$ below) in which the power-law cancels out and a universal logarithmic term remains.  

In a recent paper \cite{TanCouvreurDengJacobsen19}, Tan, Couvreur, Deng and Jacobsen  studied these correlations in 2D and 3D, confirming the exact predictions for 2D and numerically finding the exponents in 3D.  In the present paper  \cite{TanDengJacobsen20}, Tan, Deng, and Jacobsen study these correlations in 4D and 5D.  These higher dimensions are theoretically of interest because, in percolation, the upper critical dimension is 6, meaning that the behavior becomes mean-field, similar to that on the complete graph or the Bethe lattice, where exponents are known exactly and simple rational values.  A great deal of work has been done over the years in finding series expansions about six dimensions, in powers of $\epsilon = 6 - d$, and finding results in 4 and 5 dimensions is particularly useful in evaluating the accuracy of those series expansions.   In general, substantially less work has been done in percolation in 4 and 5 dimensions compared with lower dimensions.  One of the problems is the constraints on lattice size that one can simulate on a computer, considering the value of $L^d$ for $d = 4$ or 5.

Here, the authors look at correlations between two separated sets of neighboring points belonging to different combinations of clusters.    They consider $N$ neighboring points ($N = 2$ and 3) and another $N$ neighboring points separated a distance $L/2$ from the first points, in a periodic system.  They use a
a variant of the Hoshen-Kopelman algorithm on an $L^{d-1}\times L'$ lattice where $L' \gg L$ to study these correlations, joining two separate systems together to form a layer in a bulk system to study the correlations, and then returning to the un-joined systems to add additional layers.  They use a union-find algorithm \cite{NewmanZiff00} with a simple compression step and a flattening step to efficiently carry out this algorithm.  One of the advantages of using this algorithm is that they could simulate a larger $L$ than if the system were simply periodic of size $L^d$, because only one plane needs to be stored in the $L'$ direction.  

The authors look at configurations where the $N$ points in each of the two groups belong to different clusters--already something of low probability---and then look at various combinations of connectivites between the two sets of points along clusters joining them.  Finally, by looking at certain symmetric $P_\mathrm{Ns}$, asymmetric $P_\mathrm{Na}$, and mixed $P_\mathrm{Nm}$ combinations, they find quantities that also decay as a power-law in $r$, defined as the distance between these two sets of point.   This defines various exponents according to
\be
P_{\mathrm{No}} \sim r^{-2X_{\mathrm{No}}}
\ee
where o = s, m, a and $N = 2, 3$.  In their paper, they present precise values of these exponents $X_{\mathrm{No}}$.  These are new universal quantities that have never been measured before.
Because of the low probability of these events, extensive simulations were necessary to reveal the behavior.

For example, for $N = 2$, they define 
\bea
P_\mathrm{2s} &=& P(r_1, r_3 | r_2, r_4) + P(r_1, r_4 | r_2, r_3) \cr
P_\mathrm{2a} &=& P(r_1, r_3 | r_2, r_4) - P(r_1, r_4 | r_2, r_3) 
\eea
where the points $r_1,r_2$ are in one neighboring set, and $r_3,r_4$ are in another neighboring set a distance $L/2$ away from the first set.  Here the vertical bar separates the points that are connected together.  In \cite{TanDengJacobsen20}, a graphical representation of these quantities is used.  

Likewise for $N = 3$, where there are 6 total points, three function $P_\mathrm{3a}$, $P_\mathrm{3s}$, and $P_\mathrm{3m}$ are defined.
They can also extend this to $N=1$, which is identical to \eq{P2} above with $X_\mathrm{1s}=d-d_f$.  Here they quote values of $X_\mathrm{2s}$ from a recent paper of one of the authors, Ref.\ \cite{ZhangHouFangHuDeng20}
                                                                                                                                                                                                                                                                                                                                                                                                                                                                                                                                                                                                                                                                                                                                                                                                                                                                                                                                                                                                                                                                                                                                                                 
They also study the combination of correlations for $N = 2$ that is expected to show a logarithmic behavior with a universal coefficient $\delta$:
\be
F(r) = \frac{\mathbb{P}_0(r)+\mathbb{P}_1(r) -(\mathbb{P}_\mathrm{\ne})^2}{\mathbb{P}_\mathrm{2s}(r)} \sim \delta \ln r
\ee
where $\mathbb{P}_0$ is the probability that the four points all belong to different clusters, 
$\mathbb{P}_1$ is the probability that the points belong to three different clusters, with one cluster linking one point in one set to a point in the other,  and $\mathbb{P}_{\ne}$ is the probability that the two points in a single set belong to different clusters.  In 2D, we know that on a square lattice $\mathbb{P}_{\ne} = 1/4$ \cite{HuBloteZiffDeng14}, but it is not known in higher dimensions.  Carrying out precise studies, the authors find $\delta = 1.16(1)$ in 4D and 0.74(6) in 5D.   These are new results, and together with their previous work in 2D and 3D, paint a very complete picture on the universal behavior of these quantities.

The exponent $X_\mathrm{2s}$ is related to the correlation-length exponent $\nu$ by $X_\mathrm{2s} = d-y_t=d-1/\nu$, and their measurements of this exponent yields substantially increased precision of the value of this important exponent in 4D and 5D.
They find $y_t = 1.4610(12)$ or $\nu = 0.6845(6)$ in 4D and $y_t = 1.737(2) $ or $\nu =  0.5757(6)$ in 5D, compared to other recent results $\nu = 0.6852(28)$ by Koza and Po\l a \cite{KozaPola16} and 0.6845(23) by Zhang et al.\ \cite{ZhangHouFangHuDeng20}, and consistent with older results 0.6782(50) of Adler et al.\ \cite{AdlerMeirAharonyHarris90} and 0.689(10) of Ballesteros et al.\ \cite{BallesterosEtAl97} in 4D, and  $\nu = 0.5723(18)$ \cite{KozaPola16} and 0.5737(33) \cite{ZhangHouFangHuDeng20} in recent works, and 0.571(3) in Adler et al.\ \cite{AdlerMeirAharonyHarris90} from 1990 in 5D.   These various results are generally consistent with each other within error bars, despite the variety of methods that were used, highlighting the high level of understanding of percolation that has been developed over the last several decades.

Besides Monte Carlo work, another classical approach has been through series expansions about 6D, and recently Gracey \cite{Gracey15} has extended those series to higher order and has made the following predictions: $\nu = 0.6920$ in 4D and 0.5746 in 5D.  These results are a bit lower and outside the error bars of results found here, but not far off.  Perhaps higher-order series and/or future Monte Carlo work will show closer agreement. % Likewise, for $d_f$, Gracey gives $d_f = 3.0479$ in 4D and 3.528 in 5D.

For future work, it might be nice to demonstrate that these results are indeed universal by looking at another system, such as site percolation on a hypercubic lattice or site or bond percolation on a generalized FCC lattice, for example, although it is generally expected that universality should hold.  Other combinations of correlations, such as the ratio of \eq{R}, might also be of interest in higher dimensions.

\bibliographystyle{unsrt}
\bibliography{bibliography.bib}
\end{document}